\documentstyle[12pt]{article}
\newcommand{\bra}[1]{\left\langle #1\right|}
\newcommand{\ket}[1]{\left| #1\right\rangle}
\def\vec#1{\mbox{\boldmath$#1$}}
\def\mathcal{{\cal}}
\newcommand{\etal}{{\it et al.}}
\def\mathrm{\rm}
\def\d{\,\mathrm{d}}
\def\half{{\textstyle {1\over2}}}

\begin{document}
\baselineskip 24pt
\begin{center}
{\Large\bf Spin-dependent Polarizability of Nucleon with
Dispersion Relation in the Skyrme Model}
\end{center}

\baselineskip 19pt
\begin{center}
\vspace{1.5cm}
{\large Y. Tanushi\footnote
{E-mail: tanushi@nuc-th.phys.nagoya.ac.jp},
Y. Nakahara\footnote
{E-mail: nakahara@nuc-th.phys.nagoya.ac.jp},
S. Saito\footnote
{E-mail: saito@nuc-th.phys.nagoya.ac.jp}}

\vspace{.2cm}
{\it Department of Physics, Nagoya University,
Nagoya 464-01, Japan}

\vspace{.5cm}
{\large M. Uehara\footnote
{E-mail: ueharam@cc.saga-u.ac.jp}}

\vspace{.2cm}
{\it Department of Physics, Saga University,
Saga 840, Japan}

\vspace{1.5cm}
{\bf Abstract}
\end{center}

We calculate the spin-dependent polarizability of the nucleon in 
the Skyrme model. The result is compared with that of a heavy 
baryon chiral perturbation theory(HBChPT), and is shown to be the 
same as that of HBChPT up to the $\Delta$-pole terms in the 
narrow width limit of the $\Delta$ state and with the experimental 
physical constants. The effect of the $\Delta+\pi$ channel is 
rather small and is numerically quite similar to that of the 
$\Delta$ loop in the HBChPT. The electric and magnetic 
polarizabilities are recalculated using the transverse photon 
and a consistent inclusion of the $\Delta$ width.

\newpage
The electromagnetic polarizabilities are important quantities
which show the response of internal structure of nucleons to 
the external electromagnetic fields. These polarizabilities are 
extracted from the forward Compton scattering at threshold, 
and recently attracted a great deal of attention of both 
experimental and theoretical interest.

The Skyrme-soliton model is a QCD motivated model based on
the idea of large $N_c$ and of the spontaneous breaking of 
chiral symmetry. The electromagnetic polarizability is 
considered to be sensitive to the pion cloud around the 
nucleon, so that the Skyrme model may be well-suited to
the study of the polarizability.
In a previous paper{}\cite{Saito95} we have calculated the 
electric and magnetic polarizabilities in the model using 
the dispersion formula with the photo-absorption cross 
sections of the longitudinal and transversal photons for the 
electric and magnetic ones, respectively. Calculations of the
electric polarizability from the seagull term were also shown 
to be not compatible with the gauge invariance.
It was shown that the chiral leading order terms of the
electric and magnetic polarizabilities are exactly the same as 
those in the chiral perturbation theory.
\cite{Lvov93,Saito95}
Further, we have stressed that the $\Delta$ states play
important roles, and that the $N$ and the $\Delta$ states are
treated as the same rotational levels of the Skyrme soliton.

In this paper we apply the method to the study of the spin-dependent
polarizability $\gamma$ of the  nucleon.
The multipole analysis shows that $\gamma$ is small but negative:
$-1.3\,(-0.4)\times 10^{-4}\,{\mathrm{fm}}^{4}$
for the proton (neutron). \cite{Sandorfi}
On the other hand,
the chiral leading order contribution is largely positive
and is about $4.6\times 10^{-4} {\mathrm{fm}}^{4}$.
There are some studies of this quantity in terms
of the heavy baryon chiral perturbation theory(HBChPT). 
Bernard \etal{} \cite{Bernard92}
obtained $\gamma$ with the one-loop result and with including
the effect of the $\Delta$ state.
Hemmert \etal{} \cite{Hemmert97} calculated $\gamma$ up to
$O(\epsilon^3)$ with the HBChPT and the explicit
degree of freedom of the $\Delta$ state.
In this approach the small parameter $\epsilon$ is taken to be
either of the soft momentum, the pion mass,
or the mass splitting $\Delta M=M_\Delta-M_N$
between the $N$ and $\Delta$ states.
These calculations show that the contribution of the $\Delta$ state
is very large and negative. In this meaning it is very interesting
to study the spin polarizability in the Skyrme model,
since it treats the $\Delta$ state as the equal partner
as the $N$ state.

The forward Compton scattering amplitude of the nucleon
is represented as
\begin{equation}
f_1(\omega)\,\vec{\epsilon}\cdot\vec{\epsilon'}
+if_2(\omega)\omega\,\vec{\sigma}\cdot\vec{\epsilon'}
\times\vec{\epsilon},
\end{equation}
where $f_1(\omega)$ and $f_2(\omega)$ are expanded at low energies as
\begin{equation}
f_1(\omega)=-{e^2\over4\pi M}+(\bar\alpha+\bar\beta)\,\omega^2+\cdots,
\end{equation}
and
\begin{equation}
f_2(\omega)=-{e^2\kappa^2\over8\pi M^2}+\gamma\,\omega^2+\cdots .
\end{equation}
Here, $\bar\alpha$ and $\bar\beta$ are the electric and magnetic
polarizabilities, respectively, $\kappa$ the anomalous magnetic
moment of the nucleon, and $\gamma$ the
spin-dependent polarizability. \cite{GellMann,Low}
The once-subtracted dispersion relation gives for the spin-dependent
polarizability
\begin{equation}
\gamma={1\over4\pi^2}\int_{\omega_0}^\infty
{\sigma_{1/2}-\sigma_{3/2}\over \omega^3} \d\omega, \label{eq:dispgam}
\end{equation}
where $\sigma_\lambda$ denotes the photo-absorption cross section
with the helicity $\lambda$.

We calculate the photo-absorption cross section in terms of the
$\gamma+N\rightarrow \pi+N$ and $\gamma+N\rightarrow \pi+\Delta$
amplitudes in the Skyrme model. We have shown \cite{Saito95}
that the electric and magnetic Born amplitudes satisfies
the low-energy theorem of the pion photo-production amplitudes
at threshold except for the order $(m_\pi/M)^2$ term which
was recently introduced by the effect of chiral loops.\cite{Kaiser92}
The electric Born amplitude for the $\gamma+N\rightarrow \pi+N$
is given by
\begin{equation}
\left(T_E^{(-)}\right)^{N}= \left( {eG_{NN\pi}\over8\pi M} \right)
\left\{i\vec{\sigma}\cdot\vec{\epsilon}
+2{i\vec{\sigma}\cdot(\vec k-\vec{ q})(\vec{\epsilon}\cdot\vec{ q})
\over m_\pi^2+(\vec{ k}-\vec{ q})^2}\right\}, \label{eq:TEN}
\end{equation}
where $\vec{k}$ and $\vec{q}$ are the incident photon and
the outgoing pion momenta, respectively, and $\vec{\epsilon}$
is the polarization vector of the incident photon.
Here, we expanded the production amplitude as
\begin{equation}
T^a=i\epsilon_{a3b}\tau^b\, T^{(-)}+\tau^a\, T^{(0)}
+\delta_{a3}\, T^{(+)}.
\end{equation}
We see that the $\left(T_E^{(-)}\right)^N$ is of $O(N_c^{1/2})$,
while the $\left(T_E^{(+,0)}\right)^N$ are of $O(N_c^{-1/2})$
and behave as $O(\omega_k)$.
Therefore, the latter amplitudes do not lead to finite results
without unitarization of them, and are neglected in the following.
The absorption cross section is calculated to be
\begin{equation}
\Delta\sigma^N_E(\omega_k)=\left({e^2G_{NN\pi}^2\over8\pi M^2}\right)
(1-v^2)\ln{1+v\over1-v},
\end{equation}
where $\Delta\sigma=\sigma_{1/2}-\sigma_{3/2}$, and $v=q/\omega_q$
is the pion velocity.
Inserting this into eq.~(\ref{eq:dispgam}) we obtain
\begin{equation}
\gamma_E^N=
\left({e^2\over4\pi}\right){G_{NN\pi}^2\over24\pi^2M^2m_\pi^2}.
\end{equation}
In terms of the Goldberger-Treiman relation, this can be seen to be
exactly the same as that of the N-loop in ChPT,
as already shown by L'vov.
\cite{Lvov93}
The $1/m_\pi^2$ dependence shows that this is the contribution from
the pion cloud, and is of the leading order in the ChPT.
Because the proton-neutron difference
depends on the amplitude $\left(T_E^{(0)}\right)^N$,
we predict only the average between them.
It is known that the prediction of the HBChPT up to chiral order
$\epsilon^3$ includes no isospin dependence, and there is
no contribution of $\left(T_E^{(-)}\right)^N$,
$\left(T_E^{(+,0)}\right)^N$.
The multipole analysis \cite{Sandorfi} shows that
$\gamma$ is possibly negative, while the above prediction is positive.
The contribution of the $\Delta$ resonance
is considered to reverse the sign.

We now examine the contributions from the magnetic Born terms
for $\gamma+N\rightarrow \pi+N$:
\begin{eqnarray}
\left(T_M^{(-)}\right)^N &=& \left( {eG_{NN\pi}\over8\pi M} \right)
\left({\mu_V\over 2M}\right)
 \left\{ -{(\vec{\sigma}\cdot\vec{ q})
        (\vec{\sigma}\cdot\vec{ s})\over\omega_k}
-{(\vec{\sigma}\cdot\vec{ s})
  (\vec{\sigma}\cdot\vec{ q})\over\omega_k}\right.
\nonumber \\
&&\quad{}+\left.
{1 \over 2}{\left[3\vec{ s}\cdot\vec{ q}-(\vec{\sigma}\cdot{\vec{q}})
(\vec{\sigma}\cdot\vec{ s})\right]
\over\omega_k-\Delta M}
+{1 \over 2}{\left[3\vec{ s}\cdot\vec{ q}-(\vec{\sigma}\cdot\vec{ s})
(\vec{\sigma}\cdot\vec{ q})\right]
\over\omega_k+\Delta M}
\right\},
\end{eqnarray}
\begin{eqnarray}
\left(T_M^{(+)}\right)^N
&=& \left( {eG_{NN\pi}\over8\pi M} \right)
\left({\mu_V\over 2M}\right)
 \left\{ -{(\vec{\sigma}\cdot\vec{ q})
 (\vec{\sigma}\cdot\vec{ s})\over\omega_k}
+{(\vec{\sigma}\cdot\vec{ s})(\vec{\sigma}\cdot\vec{ q})
\over\omega_k}\right. \nonumber \\
&&\quad{}-\left.
{\left[3\vec{ s}\cdot\vec{ q}-(\vec{\sigma}\cdot\vec{ q})
(\vec{\sigma}\cdot\vec{ s})\right]
\over\omega_k-\Delta M}
+{\left[3\vec{s}\cdot\vec{ q}-(\vec{\sigma}\cdot\vec{ s})
(\vec{\sigma}\cdot\vec{ q})\right]
\over\omega_k+\Delta M}
\right\},
\end{eqnarray}
where $\vec{s}=\vec{\epsilon}\times\vec{k}$,
and $\mu_V$ the vector part of the nucleon magnetic moments defined
by $(\mu_p-\mu_n)/2$ in units of the nuclear magneton. 
Note that we introduced the nucleon- and $\Delta$-pole terms, and that
we have used the relation $\mu_V^{\Delta N}=-{3\over\sqrt{2}}\mu_V$
in the Skyrme model. We also see that
$\left(T_M^{(\pm)}\right)^N$ reduces to $O(N_c^{1/2})$
by the cancellation among the $N$- and $\Delta$-pole terms.
The term $\left(T_M^{(0)}\right)^N$ is of $O(N_c^{-1/2})$ and
is neglected in the following.
We rewrite $\left(T_M^{(\pm)}\right)^N$ as
\begin{equation}
\left(T_M^{(\pm)}\right)^N=\left( {eG_{NN\pi}\over8\pi M} \right)
 \left( {\mu_V\over 2M} \right)
\left\{ t_1^{(\pm)} P_1( \hat{\vec{q}}, \hat{\vec{s}})+
t_3^{(\pm)} P_3(\hat{\vec{q}}, \hat{\vec{s}}) \right\},
\end{equation}
where $P_1(\hat{\vec{q}},\hat{\vec{s}})
=(\vec{\sigma}\cdot\hat{\vec{q}})(\vec{\sigma}\cdot\hat{\vec{s}})$ and
$P_3(\hat{\vec{q}},\hat{\vec{s}})=3(\hat{\vec{q}}\cdot\hat{\vec{s}})-
(\vec{\sigma}\cdot\hat{\vec{q}})(\vec{\sigma}\cdot\hat{\vec{s}})$
are the $P$-wave projection operators for $J=1/2$ and $J=3/2$,
respectively, and $\hat{\vec{q}}=\vec{q}/q$ and
$\hat{\vec{s}}=\vec{s}/k$.
We obtain
\begin{eqnarray}
t_1^{(-)} &=&
-{1\over3M}{\Delta M q\over \omega_k+\Delta M} \nonumber\\
t_3^{(-)} &=& {1\over2M}q\omega_k
\left[{\Delta M \over \omega_k^2-\Delta M^2+i\Delta M \Gamma_\Delta}
-{2\over3}{\Delta M \over \omega_k(\omega_k+\Delta M)}\right]
\nonumber\\
t_1^{(+)} &=&
-{2\over3M}{\Delta M q\over \omega_k+\Delta M} \nonumber\\
t_3^{(+)} &=& {1\over2M}q\omega_k
\left[-{2\Delta M \over \omega_k^2-\Delta M^2+i\Delta M \Gamma_\Delta}
+{2\over3}{\Delta M \over \omega_k(\omega_k+\Delta M)}\right].
\label{eq:deltapole}
\end{eqnarray}
Here, we introduced the finite width of the $\Delta$ state by
\begin{equation}
\Gamma_\Delta={1\over6\pi}\left({G_{\Delta N\pi}\over 2M}\right)^2 q^3
\label{eq:width}
\end{equation}
with $G_{\Delta N\pi}=-(3/\sqrt{2})G_{NN\pi}$. This is the expression
given by Kokkedee without the relativistic correction.\cite{Adkins}
It gives $145\mathrm{MeV}$ with the experimental value of $G_{NN\pi}$
at $q=227$MeV.
Then, the contribution from the magnetic part to the difference of the
absorption cross section is given by
\begin{eqnarray}
\Delta\sigma_M^N &=& 8\pi\left({q\over\omega_k}\right)
\left({G_{NN\pi}\over4\pi}\right)^2\left({e\mu_V\over2M}\right)^2
\left\{2\left(|t_1^{(-)}|^2-|t_3^{(-)}|^2\right)
+\left(|t_1^{(+)}|^2-|t_3^{(+)}|^2\right)\right\}\nonumber\\
&=&
{e^2\mu_V^2\Delta M^2\over6M^2}
\Gamma_\Delta\omega_k
\times \left\{
{4\over3}{1\over \omega_k^2(\omega_k+\Delta M)^2} \right. \nonumber\\
&&\quad\quad \left.
+{16\over3}{\omega_k^2-\Delta M^2\over \omega_k(\omega_k+\Delta M)
 [(\omega_k^2-\Delta M^2)^2+\Delta M^2\Gamma_\Delta^2]}
 \right. \nonumber\\
&&\quad\quad \left.
-{6\over (\omega_k^2-\Delta M^2)^2+\Delta M^2\Gamma_\Delta^2}\right\}.
\label{eq:sigmaNM}
\end{eqnarray}
In the narrow width limit this gives for the gamma
\begin{equation}
\gamma_M^N\left|_{\Gamma_\Delta=0\, \mathrm{limit}}\right.=
-\left({e^2\over4\pi}\right){\mu_V^2\over 2M^2}
{1\over \Delta M^2}.
\end{equation}
Identifying $b_1=\mu_V^{\Delta N}/2$ we find that this is just the
$\Delta$-pole contribution in the HBChPT.\cite{Hemmert97}
$b_1$ is the constant of $O(\epsilon^2)$ counter term in the HBChPT,
and numerically about $-2.5\pm0.35$, while $\mu_V^{\Delta N}/2$ is
$-2.5$ with experimental values for the constants.
The calculated $\gamma$ in the narrow width limit is
$-4.0\times 10^{-4}\,\mathrm{fm}^4$, while we obtain
$-2.5\times 10^{-4}\,\mathrm{fm}^4$ with the finite width.
In the previous paper \cite{Saito95} the width of the $\Delta$ state
in the direct $\Delta$ pole was proportional to $v^3$ with $v$ the
pion velocity,
and was chosen so as to reproduce the experimental one.
However, this is not consistent with the width which appears
naturally in the numerator as shown in the second line of
eq.~(\ref{eq:sigmaNM}), so that the narrow width limit
does not lead to the $\Delta$-pole term in the HBChPT.
When we use the previous expression for the amplitudes we obtain
$\gamma_M^N=-6.1\times 10^{-4}\,\mathrm{fm}^4$,
which is too large compared with the case of the narrow width limit.

The interference term between the electric and magnetic terms is
given by
\begin{eqnarray}
\Delta\sigma_{\mathrm{EM}}^{N} &=&
{q\over\omega_k}{e^2\mu_V G_{NN\pi}^2\over 4\pi M^2}
\left\{ 2v {\mathrm{Re}}t_3^{(-)}
\right. \nonumber\\
&&\qquad \left.+\left[{1\over v}+ {1\over2}( 1-{1\over v^2})
\ln{{1+v\over 1-v}}
 \right]({\mathrm{Re}}t_1^{(-)}-{\mathrm{Re}}t_3^{(-)}) \right\}.
\end{eqnarray}

The Born terms for the process $\gamma+N\rightarrow \pi+\Delta$
have been also calculated in the previous  paper \cite{Saito95}:
The amplitude is expanded as
\begin{equation}
T^a=i\epsilon_{a3b}\mathcal{T}^b\, T^{(-)}+\mathcal{T}^a\, T^{(0)}
+\mathcal{T}_{a3}^+\, T^{(+)},
\end{equation}
where ${\mathcal{T}}^a$ is the transition isospin matrix from $N$ to
$\Delta$, and ${\mathcal{T}}_{a3}^+={\mathcal{T}}^a\half
\tau^3+\half{\mathcal{T}}^3_{\Delta\Delta}{\mathcal{T}}^a$.
The electric part is obtained by replacing
$\vec{\sigma}$ and $G_{NN\pi}$ in eq.~(\ref{eq:TEN})
by the transition spin operator $\vec{S}_{\Delta N}$ and
$G_{\Delta N\pi}$, respectively.
The magnetic part is given by
\begin{eqnarray}
\left(T_M^{(-)}\right)^\Delta &=&
\left( {eG_{\Delta N\pi}\over8\pi M} \right)
\left({\mu_V\over2M}\right)
 \left\{ -{(\vec{ S}_{\Delta N}\cdot\vec{ q})
        (\vec{\sigma}\cdot{\vec{s}})\over\omega_k}
-{4\over5}{(\vec{ S}_{\Delta\Delta}\cdot\vec{ q})
(\vec{ S}_{\Delta N}\cdot\vec{ s})\over\omega_q}\right. \nonumber\\
&&\quad{}+\left.
2{(\vec{ S}_{\Delta N}\cdot\vec{s})(\vec{\sigma}\cdot\vec{q})
  \over\omega_q}
-{1\over5}{(\vec{S}_{\Delta\Delta}\cdot\vec{s})
           (\vec{S}_{\Delta N}\cdot\vec{q})\over\omega_k}
\right\},
\end{eqnarray}
\begin{eqnarray}
\left(T_M^{(+)}\right)^\Delta &=&
\left( {eG_{\Delta N\pi}\over8\pi M} \right)
\left({\mu_V\over2M}\right)
 \left\{ -{(\vec{S}_{\Delta N}\cdot\vec{q})
           (\vec\sigma\cdot\vec{s})\over\omega_k}
-{1\over5}{(\vec{S}_{\Delta\Delta}\cdot\vec{q})
(\vec{S}_{\Delta N}\cdot\vec{s})\over\omega_q}\right. \nonumber\\
&&\quad{}+\left.
{(\vec{S}_{\Delta N}\cdot\vec{s})
 (\vec{\sigma}\cdot\vec{q})\over\omega_q}
+{1\over5}{(\vec{S}_{\Delta\Delta}\cdot\vec{s})
           (\vec{S}_{\Delta N}\cdot\vec{q})\over\omega_k}
\right\}.
\end{eqnarray}
The cross section for the process is given by
the electric and magnetic terms and their interference term:
\begin{equation}
\Delta\sigma^\Delta=\Delta\sigma_{\mathrm{E}}^\Delta+\Delta
\sigma_{\mathrm{M}}^\Delta
+\Delta\sigma_{\mathrm{EM}}^\Delta
\end{equation}
with
\begin{eqnarray}
\Delta\sigma_E^{\Delta}&=&
{e^2 G_{NN\pi}^2\over 4\pi M^2}{v\over b}
\left[{a-b^2\over b^2}-{a^2-b^2v^2\over2b^3v}\ln {a+b v\over a-b v}
\right],  \nonumber\\
\Delta\sigma_M^{\Delta}&=&
{e^2G_{NN\pi}^2\mu_V^2\Delta M^2
\over 24\pi M^4}{v^3\over b},\nonumber\\
\Delta\sigma_{EM}^{\Delta}&=&
-{e^2G_{NN\pi}^2{\mu_V}\Delta M\over 8\pi M^3}
{v^2\over b^3} \nonumber\\
&&\qquad\quad \times\left[ {2bv\over3}-{a^2\over bv}
+{a(a^2-b^2v^2)\over2b^2v^2}
\ln{a+bv\over a-bv} \right],
\end{eqnarray}
where $b=1+d\sqrt{1-v^2}$ and $a=(1+b^2)/2$ with $d=\Delta M/m_\pi$.

In Table 1,
we give numerical results of the spin-dependent polarizability
$\gamma$
for parameter sets I, II and III. 
Set I is that of Adkins.\cite{Adkins}
In  Set II $f_\pi$ is the experimental one
and the Skyrme parameter $e=4.0$ is chosen for $g_A$ to be reproduced.
In Set III all the constants such as $G_{\pi NN}$,
$\mu_V$ in the above are taken to be the empirical values.
Here, $\gamma=\gamma^N+\gamma^\Delta$ with
$\gamma^{N,\Delta}=\gamma^{N,\Delta}_E
+\gamma^{N,\Delta}_M+\gamma^{N,\Delta}_{EM}$.
The suffices $E$, $M$ and $EM$ denote the contributions from
the electric, magnetic terms and their interference terms,
respectively.

\begin{table}
\caption{Spin-dependent polarizability.
Those with the suffices $E$, $M$ and $EM$ are from the electric,
magnetic and interference terms between the electric and magnetic
ones, respectively.
The superscripts $N$ and $\Delta$ denote the contributions from the
$N+\pi$ and $\Delta+\pi$ channels, respectively.
$\gamma^N=\gamma_E^N+\gamma_M^N+\gamma_{EM}^N$,
$\gamma^\Delta=\gamma_E^\Delta+\gamma_M^\Delta+\gamma_{EM}^\Delta$,
and $\gamma=\gamma^N+\gamma^\Delta$.
All values are in units of $10^{-4}\, \mathrm{fm}^{4}$.
See text for the parameter sets.
}
\begin{center}
\begin{tabular}{cccccccccc}
\hline
Set & $\gamma_E^N$ & $\gamma_M^N$ & $\gamma_{EM}^N$ & $\gamma^N$ &
$\gamma_E^\Delta$ & $\gamma_M^\Delta$ & $\gamma_{EM}^\Delta$ &
$\gamma^\Delta$ & $\gamma$ \\ \hline
 I  & 3.9 & $-1.2$ & $-1.3$ & 1.4 &
     $-0.3$ & 0.0 & 0.0 & $-0.3$ & 1.1 \\
 II & 2.9 & $-2.8$ & $-0.8$ & $-0.7$ &
     $-0.3$  & 0.0 & 0.0 & $-0.3$ & $-0.9$ \\
 III & 5.0 & $-2.5$ & $-2.4$ & 0.2 &
     $-0.4$ & 0.1 & 0.0 & $-0.3$ & $-0.1$  \\ \hline
\end{tabular}
\end{center}
\end{table}

The result of Set III can be compared with that of the HBChPT.
Hemmert \etal\, showed in ref. \cite{Hemmert97} that
$\gamma$ is given by $[4.5$ ($N$ loop) $-4.0$ ($\Delta$ pole) $-0.4$
($\Delta$ loop)]$\times 10^{-4}\,\mathrm{fm}^4$.
As already shown, $\gamma_E^N$ is the same as the $N$-loop term in the
HBChPT. Note that we did not use the Goldberger-Treiman relation here.
$\gamma_M^N$ reduces to the $\Delta$-pole term
at the narrow width limit.
For the finite width case $\gamma^N$ is close to th sum of the
$N$-loop and $\Delta$-pole terms in the HBChPT.
The effect of the $\Delta+\pi$ channel is small and is also close to
that of the $\Delta$ loop in the HBChPT.
In the latter the contribution of magnetic terms is not included,
so that the electric term can be compared with that of $\Delta$ loop.
We see that the total result is very similar to that of the HBChPT.
Numerical results show that the magnetic terms for the $\Delta+\pi$
channel are small but have an opposite sign.

Here, we discuss about the Drell-Hearn-Gerasimov(DHG)
sum rule\cite{DHG} in the Skyrme model.
The low-energy limit of the Compton scattering amplitude
is given by the nucleon Born terms with the spatial component of
the electromagnetic current, and we obtain
\begin{eqnarray}
T_{\mathrm{N-pole}}&=&{1\over4\pi}\left\{
{ \bra{N(\vec{p})}\vec{\epsilon}'\cdot\vec{J}_{em}
  \ket{N(\vec{p}+\vec{k})}
  \bra{N(\vec{p}+\vec{k})}\vec{\epsilon}\cdot\vec{J}_{em}
  \ket{N(\vec{p})}
  \over -\omega_k} \right. \nonumber \\
&&\, + \left.
{\bra{N(\vec{p})}\vec{\epsilon}\cdot\vec{J}_{em}
 \ket{N(\vec{p}-\vec{k})}
 \bra{N(\vec{p}-\vec{k})}\vec{\epsilon}'\cdot\vec{J}_{em}
 \ket{N(\vec{p})}
\over \omega_k}\right\}.
\end{eqnarray}
The spin-dependent part of the $N$-pole terms is then given by
\begin{equation} T_{\mathrm{N-pole}} =
-{e^2\over8\pi}{{\mu_N}^2\over M_N^2}\,\omega_k\,
 i\vec{\sigma}\cdot\vec{\epsilon}'\times\vec{\epsilon},
\end{equation}
where $\mu_N$ is the nucleon magnetic moment in units of
nuclear Bohr magneton.
Consequently, the unsubtracted dispersion relation is
\begin{equation}
-{e^2\mu^2_N\over8\pi M^2}={1\over4\pi^2}\int_{\omega_0}^\infty
{\sigma_{1/2}-\sigma_{3/2}\over \omega} d\omega. \label{eq:DHG}
\end{equation}
This is different from the DHG sum rule,
in which the left-hand side is given by
the anomalous magnetic moment instead of the total magnetic one.
How to resolve this has been shown by Low \cite{Low}:
The Born terms in terms of the time-component
of the relativistic electromagnetic current removes this discrepancy;
however,
the effect is highly relativistic and cannot be obtained in a
nonrelativistic approach such as in the Skyrme model.
The contribution from the electric Born terms is already calculated by
L'vov \cite{Lvov93} and is given by
\begin{equation}
{e^2G_{NN\pi}^2\over 32\pi^3 M^2},
\end{equation}
while those from the magnetic terms are calculated to be in the narrow
width approximation for the $\Delta$ state
\begin{equation}
-{e^2\mu_V^2\over 8\pi M^2}. \label{eq:delta}
\end{equation}
Noting that $\mu_N=\mu_S+\tau_3\mu_V$ with $\mu_S/2M$ and $\mu_V/2M$
to be $O(N_c^{-1})$ and $O(N_c)$, respectively, we see that
the left-hand side of eq.(\ref{eq:DHG}) is of $O(N_c^2)$,
but the right-hand side is of $O(N_c)$, so that there seems to be
an inconsistency in the $N_c$ dependence of the DHG sum rule.
However, the term of $O(N_c^2)$ in the left-hand side is completely
canceled by the contribution from the $\Delta$ resonance as shown
in eq.(\ref{eq:delta}),
and it turns out that the left-hand side of the sum rule becomes
of $O(1)$, but the right-hand side is of $O(N_c)$.
It is not clear if the sum rule is down to that of $O(1)$,
because of the nonrelativistic approach.
A similar situation was also shown in the Adler-Weisberger(AW)
relation \cite{Uehara91}:
The square of the axial-vector constant is of $O(N_c^2)$,
but other terms is at most of $O(N_c)$.
Therefore, there appears an inconsistency in the $N_c$ dependence
of the AW relation, but the contribution of the $\Delta$
states again cancels the term of $O(N_c^2)$ at the narrow width limit.
In this case the remaining terms of $O(N_c)$ reduces to be $O(1)$,
due to further cancellations among the isospin odd forward scattering
amplitudes.

\begin{table}
\caption{Electric and magnetic polarizabilities,
$\bar\alpha$ and $\bar\beta$.
Notations are the same as those of the spin-dependent polarizability
$\gamma$.
All values are in units of $10^{-4}\,\mathrm{fm}^{3}$.
See text for the parameter sets.}
\begin{center}
\begin{tabular}{ccccccccccccc}
\hline
Set  & $\bar\alpha^N$ & $\bar\alpha^\Delta$ & $\bar\alpha$
&$\bar\beta_E^N$ & $\bar\beta_M^N$ & $\bar\beta_{EM}^N$ & $\bar\beta^N$
& $\bar\beta_E^\Delta$ & $\bar\beta_M^\Delta$ & $\bar\beta_{EM}^\Delta$
& $\bar\beta^\Delta$ & $\bar\beta$  \\ \hline
 I  & 10.8 &  6.4 & 17.2 &
 1.1 & 4.3 & 0.1 & 5.5 & 0.1 & 0.5 & $-1.3$ & $-0.7$ & 4.8 \\
    &  & (5.6) & (16.4)   &   &      &  &    & (0.9)  &
    &     & (0.1)  & (5.6) \\
 II & 8.0 & 5.3 & 13.3 & 
 0.8 & 7.7 & $-0.3$ & 8.2 & 0.1 & 0.4 & $-1.1$ & $-0.5$ & 7.7 \\
    &      & (4.7)& (12.7) &    &      &      &      & (0.7) &
    &     & (0.0)  & (8.2)  \\
III & 13.9 & 8.2  & 22.1 &
 1.4 & 9.0 & 0.3 & 10.7 & 0.1 & 1.4 & $-2.4$ & $-0.9$ & 9.8  \\
    &      & (7.2)& (21.1) &      &      &      &      & (1.1) &
    &     & (0.1)  & (10.8)     \\   \hline
\end{tabular}
\end{center}
\end{table}

Finally, we calculate the spin-independent part; namely,
the electric and magnetic polarizabilities of the nucleon.
In ref. \cite{Saito95} the electric polarizability $\bar\alpha$ is
derived from the absorption cross section by the longitudinal photon,
thereby the magnetic one can be obtained by means of
the Baldin sum rule. \cite{Baldin}
However, it is possible to calculate directly them using the transverse
photon. \cite{Lvov93}
The difference between two approaches appears only in the case of
the $\Delta+\pi$ channel for the final states.
Here we give the result with use of the transverse photon.
In this calculation we also used the
amplitudes for the direct $\Delta$-pole terms to the $N+\pi$
channel given in eq.~(\ref{eq:deltapole}).
The magnetic Born term to the $\pi+N$ channel at the narrow width limit
is again the same as that of the $\Delta$-pole term in the HBChPT:
\begin{equation}
\beta_M^N\left|_{\Gamma_\Delta=0\,\mathrm{limit}}\right.
=\left({e^2\over4\pi}\right){\mu_V^2\over M^2}
{1\over\Delta M}.
\end{equation}
The numerical results for the electric and magnetic polarizabilities
are shown in Table 2.
The numbers in parentheses are values calculated
using the longitudinal photon.
Empties show no change for this case.
We can see that
the results for Set III are very close to those of the HBChPT:
Hemmert \etal\, showed  in the calculation up to $O(\epsilon^3)$ that 
$\bar\beta$ is given by $[1.2(\mbox{N loop})+12(\mbox{$\Delta$ pole})
+1.5(\mbox{$\Delta$ loop})]\times 10^{-4}\,\mathrm{fm}^3$.
\cite{Hemmert97}
The effect of the finite width is seen to reduce the value of the
$\Delta$-pole term.
$\beta_E^\Delta$ is, however, small, but the interference term
$\beta_{EM}^\Delta$ is rather large and negative.
The large interference term is due to the high-energy behavior of the
amplitudes of the magnetic part. This may need further consideration.
In the previous paper \cite{Saito95}
the calculated results for $\bar\beta$  are $7.8$ and $21.3$
in units of $10^{-4}\,\mathrm{fm}^3$ for Set I and III,
respectively.\footnote{There was an error in the expression and
the numerical results in the contribution from the $\Delta+\pi$ channel
for the magnetic polarizabilities. \cite{Saito97}}
Therefore, we see that the inconsistent inclusion of the $\Delta$ width
leads to too large values for the magnetic polarizability.
The effect of the finite width makes the magnetic polarizabilities
rather small in the consistent inclusion of this paper.

In summary we have calculated the spin polarizability of the nucleon
in the Skyrme model, where the pion photo-production Born amplitudes
are employed for obtaining the absorption cross section
in the dispersion relation.
The electric and magnetic Born terms agree with the $N$-loop and the
$\Delta$-pole terms of the HBChPT, at the limit of the narrow width
of the $\Delta$ state.
The electric and magnetic polarizabilities were also calculated using
the transverse photon and by the consistent treatment of
the $\Delta$ width.

\end{document}